# Low-Rate Smartphone Videoscopy for Microsecond Luminescence Lifetime Imaging with Machine Learning

Yan Wang[1], Sina Sadeghi[1], Rajesh Paul[1], Zach Hetzler[1], Evgeny Danilov[2], Frances S. Ligler[3], Qingshan Wei[1]*

[1] *Department of Chemical and Biomolecular Engineering, North Carolina State University, Raleigh, NC, 27695 USA.*

[2] *Department of Chemistry, North Carolina State University, Raleigh, NC, 27695 USA.*

[3] *Department of Biomedical Engineering, Texas A&M University, College Station, TX 77843, USA.*

* Corresponding email: qwei3@ncsu.edu

**Abstract:**

Time-resolved techniques have been widely used in time-gated and luminescence lifetime imaging. However, traditional time-resolved systems require expensive lab equipment such as high-speed excitation sources and detectors or complicated mechanical choppers to achieve high repetition rates. Here, we present a cost-effective and miniaturized smartphone lifetime imaging system integrated with a pulsed UV LED for 2D luminescence lifetime imaging using a videoscopy-based virtual chopper (V-chopper) mechanism combined with machine learning. The V-chopper method generates a series of time-delayed images between excitation pulses and smartphone gating so that the luminescence lifetime can be measured at each pixel using a relatively low acquisition frame rate (e.g., 30 fps) without the need for excitation synchronization. Europium (Eu) complex dyes with different luminescent lifetimes ranging from microseconds to seconds were used to demonstrate and evaluate the principle of V-chopper on a 3D-printed smartphone microscopy platform. A convolutional neural network (CNN) model was developed to automatically distinguish the gated images in different decay cycles with an accuracy of >99.5%. The current smartphone V-chopper system can detect lifetime down to ~75 µs utilizing the default phase shift between the smartphone video rate and excitation pulses and in principle can detect much shorter lifetimes by accurately programming the time delay. This V-chopper methodology has eliminated the need for the expensive and complicated instruments used in traditional time-resolved detection and can greatly expand the applications of time-resolved lifetime technologies.

**TOC: Smartphone V-Chopper for Luminescence Lifetime Imaging**

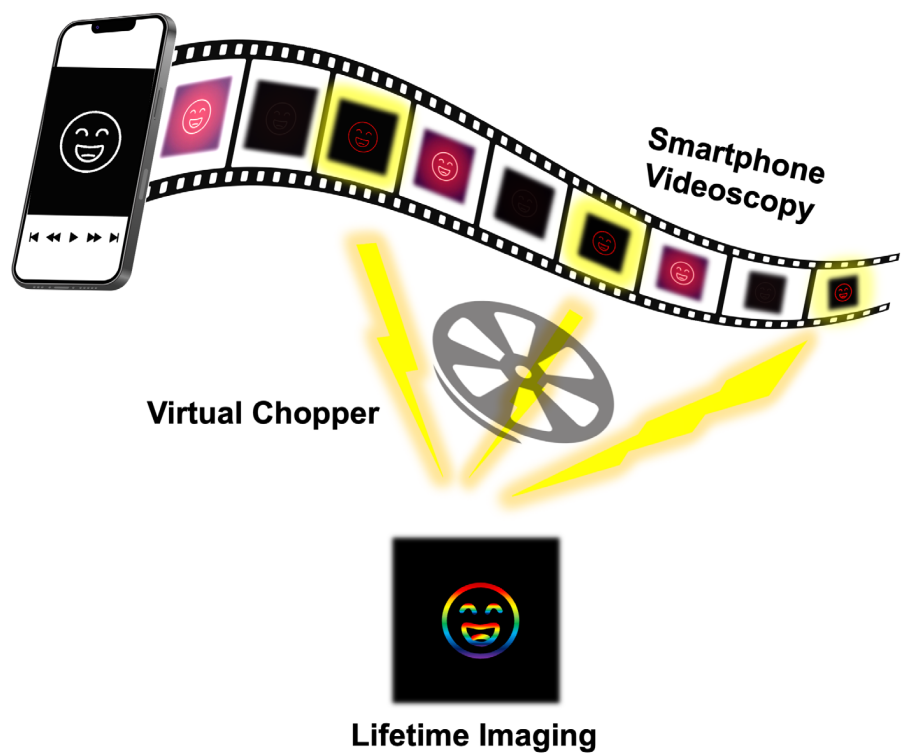

Time-resolved techniques, including time-gated autofluorescence-free imaging and fluorescence lifetime detection, have drawn significant attention in the past decades[1,2]. By taking advantage of a long-lived luminescence probe[3–6], the high background scattering and autofluorescence in biological samples can be effectively removed using time-resolved luminescence detection. Moreover, coded luminescence lifetimes can be exploited for temporally multiplexed detection assays, enabling multi-channel detection while minimizing crosstalk between detection channels, which is a common limitation for the spectral multiplexing method[7,8]. Therefore, the time-resolved detection and luminescence lifetime imaging has found a wide variety of applications, such as high-contrast, in-vivo imaging of cells and tissues[9–11], detection of rare diseased cells and pathogenic microorganisms[12–14], ultrasensitive bioassays[15–17], and physiological sensing (e.g., pH and temperature)[18–20]. A range of analytical instruments such as spectrometers, microscopes, and flow cytometers have been adapted to enable time-resolved luminescence measurements[21–26]. However, most current systems for time-resolved and lifetime detection require complicated mechanical choppers to achieve high repetition rates or expensive equipment such as high-speed excitation sources and detectors such as photomultiplier tubes (PMT), streak cameras, and intensified CCD cameras to provide the temporal resolution [26–29]. The bulkiness and complexity of the current time-resolved instruments, therefore, have posed significant challenges to the broad access to this technology outside of well-equipped laboratories.

Portable and cost-effective time-resolved devices are promising new platforms for point-of-care monitoring for medical, agricultural, and environmental applications. In particular, a modern smartphone equipped with advanced process unit and camera modules is an emerging platform for field-portable time-gated or time-resolved detection.[30] For instance, time-gated imaging has been adopted on the smartphone by capturing persistent post-excitation luminescence[31,32]. On the other hand, lifetime resolving and imaging can be achieved by exponentially fitting the pixel intensities in consequent time-gated frames over time on the smartphone[33–38]. However, the temporal resolution of smartphone camera is very limited due to its low frame rate (typically 30 frames per second (fps)). Even though the frame rate can reach 60 fps or higher on some smartphone models, it is still inadequate to detect lifetimes in the range of sub-millisecond (ms). As such, for demonstration purposes, many previous smartphone-based time-gated platforms[31–34,38] selected persistent luminescent phosphors with ultra-long lifetimes of hundreds of milliseconds to seconds (s). Additional mechanical apparatus such as chopper and motorized turntable were necessarily included into the system to break the limitation of temporal resolution of the smartphone for the measurement of shorter lifetimes in the microsecond (μs) range.[35–37] To our knowledge, no methodology has been reported so far using a standalone smartphone system without any type of chopper or motor to detect the lifetimes down to microseconds.

Here, a cost-effective and miniaturized smartphone-based lifetime imager was developed for luminescence lifetime quantification on the microsecond time scale using a virtual chopper (V-chopper) concept combined with machine learning. The smartphone V-chopper system was integrated with a pulsed UV LED, a UV reflection mirror, and a 615 nm band-pass filter in a 3D-printed enclosure. The V-chopper mechanism used the video rate (30 fps) of the smartphone to record repeated luminescence decay cycles and a convolutional neural network (CNN) model to extract correct gated images with >99.5% accuracy from different modulation cycles for lifetime image reconstruction. The effect of light pulse frequency, duty cycle, smartphone frame rate, and exposure time were systematically studied both experimentally and theoretically. Under the optimal setting, the smartphone V-chopper system can resolve luminescence lifetime from Europium (Eu) complex dyes as short as 75 µs. This portable smartphone V-chopper system decoupled the traditional time-resolved detection from expensive and complicated instruments such as mechanical choppers and high-speed detectors, making lifetime measurements practical in resource-limited settings.

## Results

**Design of the Smartphone V-Chopper Lifetime Imaging Device**

The prototype for the smartphone V-chopper lifetime imaging device utilized a 3D-printed enclosure to integrate a UV LED (365nm, M365L3, Thorlabs), a condenser lens (ACL2520U-A, Thorlabs), and a UV-enhanced reflection mirror (PFSQ10-03-F01, Thorlabs) with the smartphone camera (**Figure 1a,b**). The UV LED was controlled by an LED driver (LEDD1B, Thorlabs) in the Trigger or Modulation Mode and pulsed by a square wave voltage source (DG1062Z, Rigol). A 615 nm band-pass filter (87-739, Edmund Optics) was inserted in the enclosure in front of the phone camera when detecting luminescent signals from Europium complex dyes. The dyes were deposited on the glass slide, which was inserted at the bottom of the 3D-printed enclosure. A smartphone (Samsung Galaxy S9) with manual video control (e.g., ISO, focal length, shutter speed, video frame rate, and image resolutions) was placed on top of the enclosure as the detector. A 60 fps video frame rate was used for lifetime detection of ultra-long luminescent materials (seconds). For the measurement of microsecond lifetime targets, a normal video rate (30 fps) was used combined with the V-chopper principle, as detailed below.

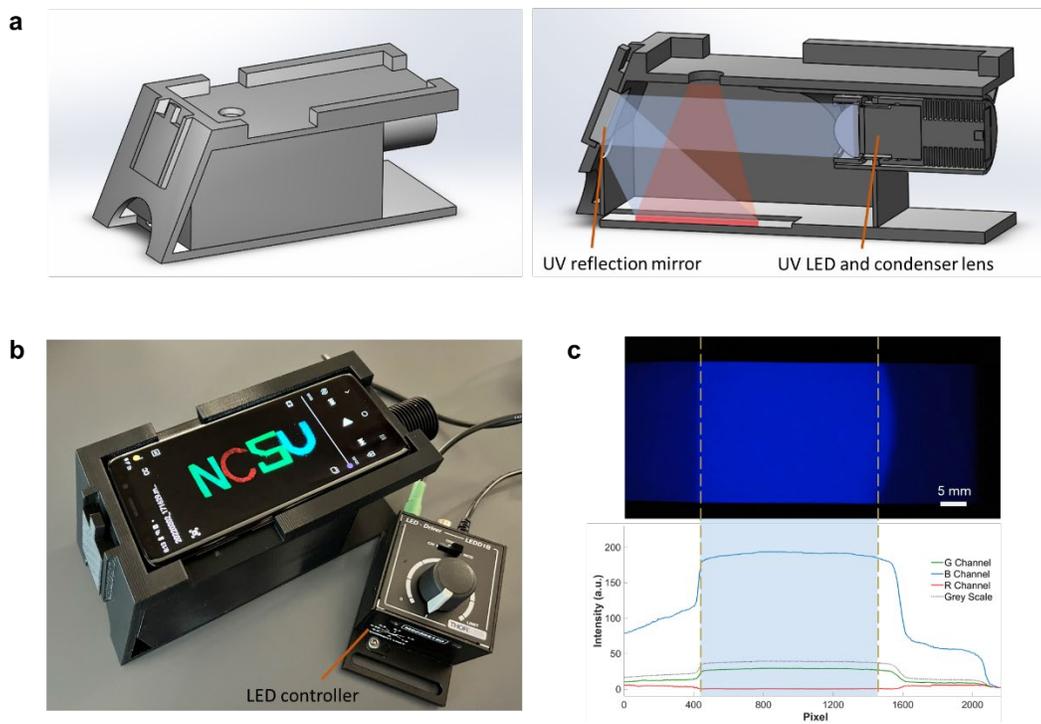

**Figure 1 | Design of the smartphone V-chopper lifetime system. a,** Schematic and cross-section of the smartphone enclosure. **b,** Photograph of the actual smartphone V-chopper system with LED and pulse control. **c,** Illumination distribution of light from the LED on the sample slide.

To deliver a uniform illumination to the sample slide, a 1" × 1" UV-enhanced mirror was used to reflect the excitation instead of direct illumination. The highly divergent emission from the UV LED was first collimated using the aspheric condenser lens (f=20.1 mm, NA=0.60) and then evenly projected on the glass slide by a tilted mirror (**Figure 1a**). The tilted angle of the mirror is designed to be ~67.5 degree relative to the slide surface, so that the illumination center is aligned with the field of view (35 × 63 mm$^2$) of the smartphone. **Figure 1c** shows an autofluorescence image of a printing paper under UV excitation. The fluorescence intensity across the slide was extracted with Matlab for R, G and B channels, respectively, showing the uniformity of the signals for subsequent measurements and experiments.

**Resolving ultra-long lifetime in single decay cycle**

We first tested the smartphone device for lifetime imaging of persistent luminescent probes by acquiring multiple gated images per decay cycle (**Figure 2a**). The ultra-long or persistent luminescent phosphors can glow for a relatively long time from seconds to even days after switching off excitation sources. The long luminescence is also called "afterglow".[39] Four afterglow composite powders of Calcium Sulfide (red) and Strontium Aluminate Europium Dysprosium (malachite, jade green, and cyan) were used as the testing samples and evenly dispensed on the adhesive tape, forming four letters "N", "C", "S", and "U", respectively (**Figure 2b**). The glowing letters were sandwiched with glass slides and then inserted into the smartphone enclosure for lifetime imaging. The letters were excited for 10 seconds (0.005 Hz and 5% duty cycle) with an irradiance of 1 mW/cm$^2$. No band-pass filter was used in this application in order to capture all four colors in the visible wavelength range. The smartphone took a video recording both UV on (10 s) and UV off (~190 s) time periods at a frame rate of 60 fps and exposure time of 1/60 s (**Supporting Video V1**). The image frames were then extracted from the recorded video by Matlab, and the time-gated frames were identified based on the background autofluorescence level. The last frame when the UV was on was assigned as the #0 frame (**Figure 2b, top**), and the first frame after UV off was labeled as the #1 frame, which is also the first time-gated image in the gating cycle (**Figure 2b, middle**). The delay time equals the integer times of the frame interval which is the reciprocal of frame rate (e.g., 1/60 s for frame #1, 2/60 s for frame #2, 3/60 s for frame #3, and so on). Totally 8,000 gated frames (from #1 to #8000) after UV off were identified per decay cycle to resolve the luminescent lifetime. The lifetime images were then reconstructed based on the lifetimes determined from each pixel, which was calculated by exponentially fitting the intensities of each pixel over the delay time using the gated frames. The bottom panel in **Figure 2b** shows a representative smartphone lifetime image generated by Matlab, where different colors represent different lifetime values (0-40 seconds). The image clearly shows different lifetime values for the different letters. For instance, letter "C" has the shortest lifetime of around 1 seconds and letter "U" has the longest lifetime of about 30 seconds. The average pixel intensities for each of four letters were plotted (solid lines) in **Figure 2d** as the luminescence decay curves, confirming the different lifetimes for different phosphors.

To validate the lifetimes of glowing letters measured on the smartphone, we used a conventional benchtop microscope (Olympus BX43) as a comparison. Two regions of interest (ROI) were selected to be imaged by a 4× objective on the benchtop system (**Figure 2b, top, yellow and red boxes**). The same pulsed LED from the smartphone device was used to excite the letters with the same irradiance of 1 mW/cm$^2$ for 10 seconds. The benchtop microscope recorded videos at 15 fps with the exposure time of 1/60 s. The #0 frame before UV off and #1 frame after UV off are shown in **Figure 2c (top and middle panels)**. The lifetime images from the benchtop microscope were calculated in the same way as mentioned above, shown in **Figure 2c (bottom panels)**. When comparing them to the lifetime image obtained from the smartphone device, the lifetime values are quite consistent for all four letters between the different lifetime images. Moreover, the normalized intensity of luminescence decay curves generated by the two acquisition systems matched with each other well (**Figure 2d**).

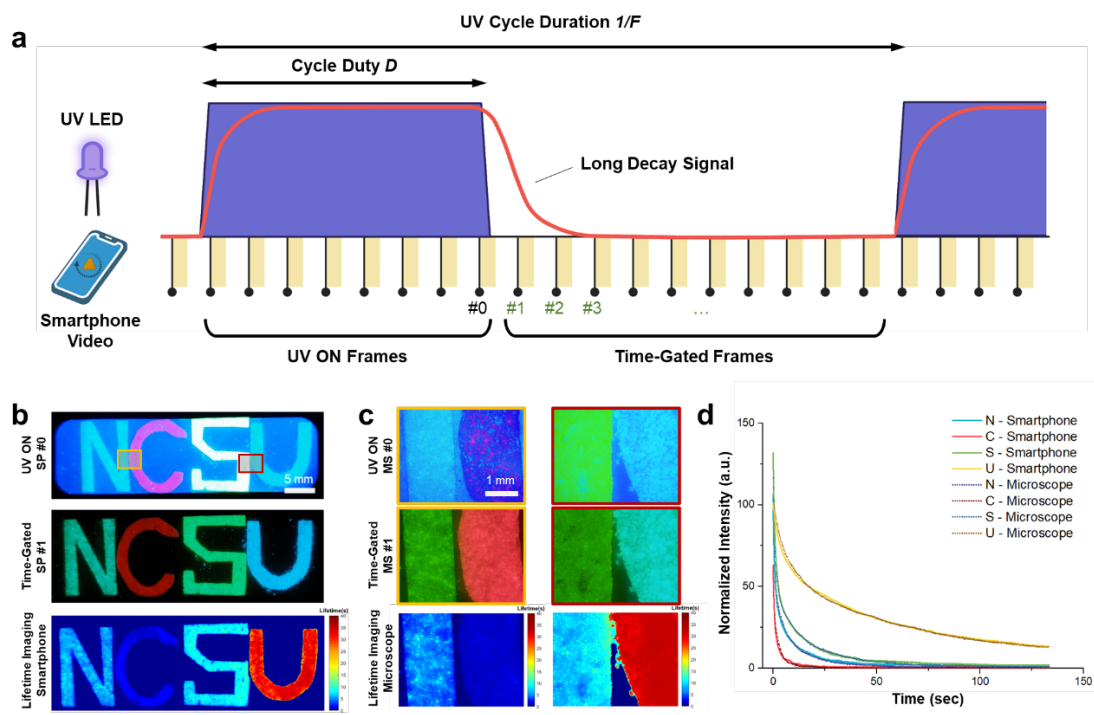

**Figure 2 | Smartphone-based lifetime imaging of ultra-long (seconds) luminescent targets. a,** Schematic of smartphone videoscopy for long lifetime measurement. Multiple gating can be applied to a single luminescence decay cycle. **b,** From top to bottom: smartphone fluorescence image (Frame #0, UV on), smartphone time-gated image (Frame #1, UV off), and smartphone lifetime image, respectively. **c,** From top to bottom: benchtop fluorescence images, benchtop time-gated images (Frame #1, UV off), and benchtop lifetime images, respectively. Two different ROIs were selected from **b (top)** for comparison. **d,** Luminescence decay curves extracted from the time-gated frames on smartphone (solid lines) and benchtop (dashed lines) microscopes, respectively.

**The V-chopper concept**

While it is easy to implement on the smartphone, the previous gating method (**Figure 2**) is limited by the intrinsic low frame rate of the smartphone, and therefore lacks the required temporal resolution to probe faster luminescent decay events in the microsecond range. To overcome this limitation, a new virtual chopper (V-chopper) method is introduced, allowing video-rate smartphone device to detect microsecond lifetime signals without the need for precision excitation synchronization. The basic principle of V-chopper concept is illustrated in **Figure 3**. The method consists of three simple steps: 1) smartphone videoscopy, to capture multiple cycles of luminescence decay driven by pulsed excitation; 2) frame extraction by machine learning, to isolate time-gated images (UV-off images) from different decay cycles and rearrange those frames to form a new virtual gated image sequence that can represent the illuminance decay property of the dye; and 3) lifetime image reconstruction, to calculate the lifetime value for each pixel and recover the 2D lifetime image. The fundamental difference between the V-chopper method and the conventional gating method is that V-chopper is based on only one image gating per or every a few decay cycles, while most conventional methods require continuous gating for each decay cycle, and therefore have a significantly higher requirement on the sampling frame rate.

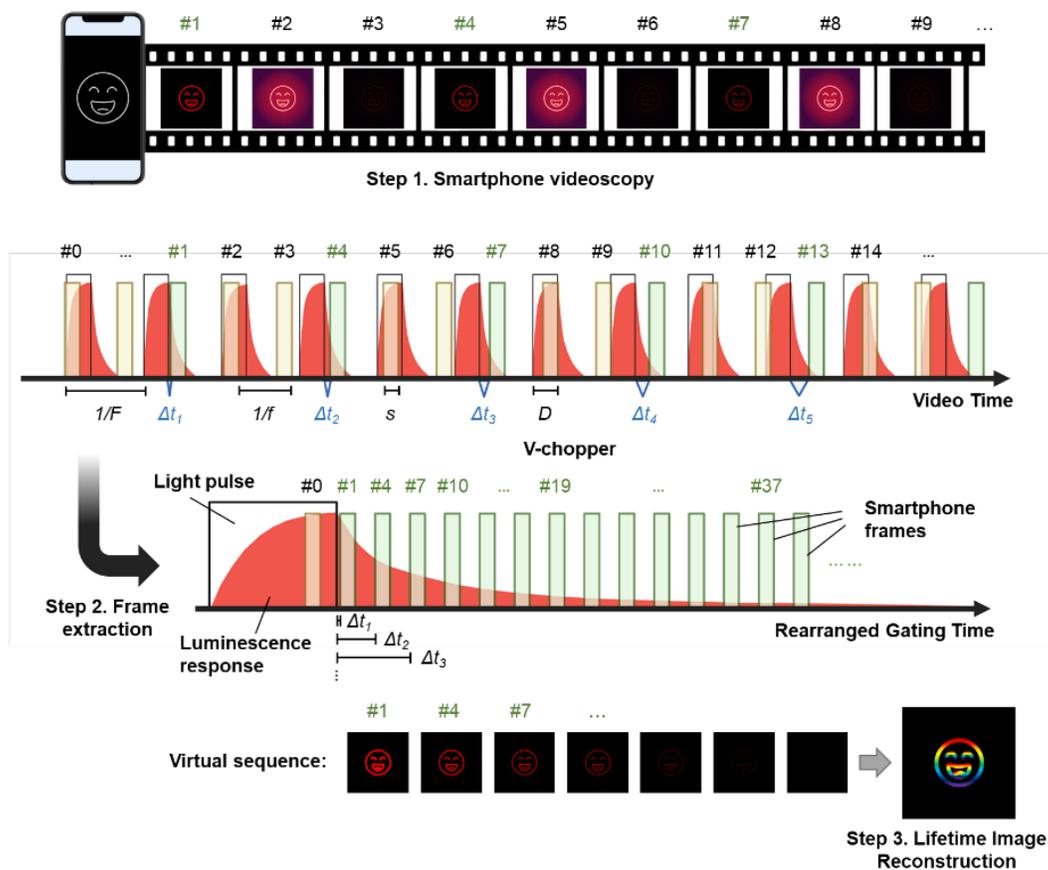

**Figure 3 | Concept of V-chopper.** The method consists of three steps: 1) smartphone videoscopy, 2) frame extraction by machine learning, and 3) lifetime image reconstruction. In Step 1, the large clear bars (black edge) indicate the timing when the UV excitation was on, while the translucent bars indicate the timing of the detection measurements. The red curves illustrate time-dependent emission intensity. In Step 2, image frames (translucent bars) without the excitation on were collected to measure the emission lifetimes.

**Machine learning-assisted gated image extraction**

To streamline the selection and rearrangement of the gated images in the V-chopper method, a convolutional neural network (CNN) model was designed to automatically discriminate images of interest with a high accuracy (**Figure 4**). For accurate lifetime determination, the raw smartphone video frames need to be classified into two different groups: UV-on images (or Class 1) and UV-off images (or Class 0). After classification, all Class 1 images will be discarded, and Class 0 images will be rearranged based on their intensity level to form the virtual time-gated image sequence. The CNN model developed here was composed of three convolution layers followed by a flattened, fully connected layer including 100 hidden nodes (**Figure 4** and **Supporting Table S1**). Each convolution layer was succeeded by batch normalization, ReLU activation, and a max pooling layer. The fully connected layer was followed by batch normalization, ReLU activation, and dropout. Lastly, Softmax activation was utilized in the output layer to generate class probabilities, resulting in predicted labels. Additional strategies were applied to combat overfitting (see **Methods**).

The CNN model was trained and tested with a balanced number of Class 1 and Class 0 smartphone images, each composed of 3,200 images. Briefly, the smartphone video frame images were converted into gray scale and resized to 128 x 128 pixels. Then, the dataset was divided into three subsets - training, validation, and test sets - with the split ratio of 60/20/20. The learning process was performed in three steps: first, the CNN model was trained on the training set and applied on the validation set; second, the model was trained on both training and validation sets and applied on the test set; Finally, the model was trained using the whole dataset. The final trained model (after all three steps) was exported for future classification use. The trained CNN model was then applied to unknown images and separated images into two categories (Class 1 or Class 0). The images with predicted label "0" were the time-gated frames (UV off) which were then used to calculate the luminescence lifetime.

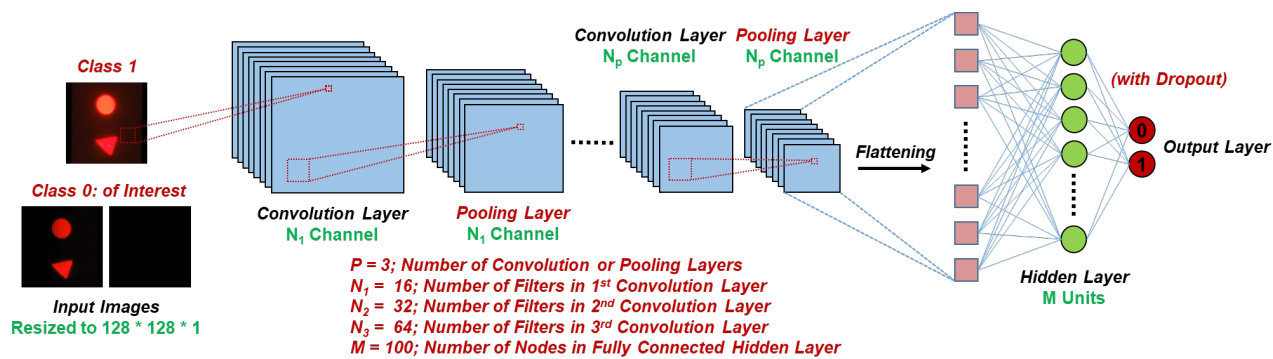

**Figure 4 | Workflow of the CNN model for automatically classifying smartphone video frames.** Class 0: UV-off images, or gated frames with no autofluorescent background; Class 1: UV-on images, or frames with autofluorescent background.

**Microsecond lifetime imaging by V-chopper and machine learning**

For a proof-of-concept demonstration of the V-chopper method, two Europium (Eu) probes with distinct lifetimes in the range of microseconds were patterned on a paper substrate in different shapes (**Figure 5, and Methods**). One Eu chelate, 4,4'-bis(1",1",1",2",2",3",3"-heptafluoro-4",6"-hexanedion-6"-yl) chlorosulfo-o-terphenyl-Eu$^{3+}$ (BHHCT-Eu$^{3+}$) with a shorter lifetime (~250 μs) was patterned in the round shape and Eu microbeads with a longer lifetime (~500 μs) were dispersed in a triangle shape. The sample slide was illuminated using the UV LED at a pulse frequency of 50 Hz and 40% duty cycle, and the phosphorescence was recorded at 30 fps and 1/350 s exposure time set on the smartphone (**Figure 5a,** and **Supporting Video V2-3**). A series of time-gated images from different modulation cycles were generated (**Figure S1b**). The frame with UV on is assigned at the #0 frame and the first frame after the UV is off is referred as the #1 frame, as usual. The time delay of #1 frame is close to 0 ($\Delta t_1 \sim 0$). Subsequently, #4, 7, 10, … are time-gated images identified by the CNN model with different decay time $\Delta t_n$. Although the pre-set frame rate was 30 fps, we noticed the actual frame rate was around 29.98 fps. The little shift of the smartphone video rate is actually critical to generate small time delays without the need for expensive control devices. The interplay between video rate and excitation frequency was explored in more detailed by a modeling method as described in the **Discussion** section. Finally, according to Equation 6, time interval $\Delta t$ is 22.2 μs between two successive frames for an actual frame rate of 29.98 fps. For a UV pulse of 50 Hz, one gated image was identified for every 3 consecutive image frames, and therefore the gated interval is $\Delta t_2 = 66.6$ μs, $\Delta t_3 = 123.2$ μs, and so on for the rearranged virtual image sequence.

The developed CNN model classifies time-gated images with a high accuracy. **Figure 5b and c** show the confusion matrix as well as the performance measures of the CNN model. The CNN model trained on the combination of training/validation sets was found to perform well with a training and a test accuracy of 99.90 and 99.84 %, respectively (**Figure 5c**). The CNN model trained on the whole data set was subsequently applied on 5317 unseen images to classify. The smartphone images were automatically separated into two different categories by their predicted labels. The representative #0 frame as well as CNN-classified time-gated images are shown in **Figure 5d**. The lifetimes of these two dyes can be easily distinguished in the gated frames. The circle is barely visible after frame #37 while the triangle still glows slightly in frame #73, indicating the much longer lifetime of triangle (Eu microbeads) than the circle (BHHCT-Eu). The average pixel intensities of the circle (blue) and triangle (red) were plotted as a function of time (**Figure 5e**). When fitted with the mono-exponential decay curves, the lifetime of the circle was calculated to be 168.4 µs, and the lifetime of the triangle was 512.3 µs. A reconstructed lifetime image was generated by fitting the intensity of each pixel over time and calculating the lifetime for each pixel (**Figure 5f**). A clear difference in the lifetimes of the circle and the triangle was visualized based on the color bar. The corresponding lifetime histogram of each pattern is shown in **Figure 5g**, showing a narrow distribution of ±9.3 µs and ±14.5 µs for the circle and the triangle, respectively. Both lifetimes determined on the smartphone V-chopper device were highly consistent with the results measured on the commercial time-resolved spectrometer, which were 167.2 ± 1.4 µs and 516.4 ± 7.0 µs for the circle and the triangle, respectively (**Figure S2 and Table S2**). The error percent is within 0.8% when comparing the smartphone V-chopper and benchtop spectrometer results.

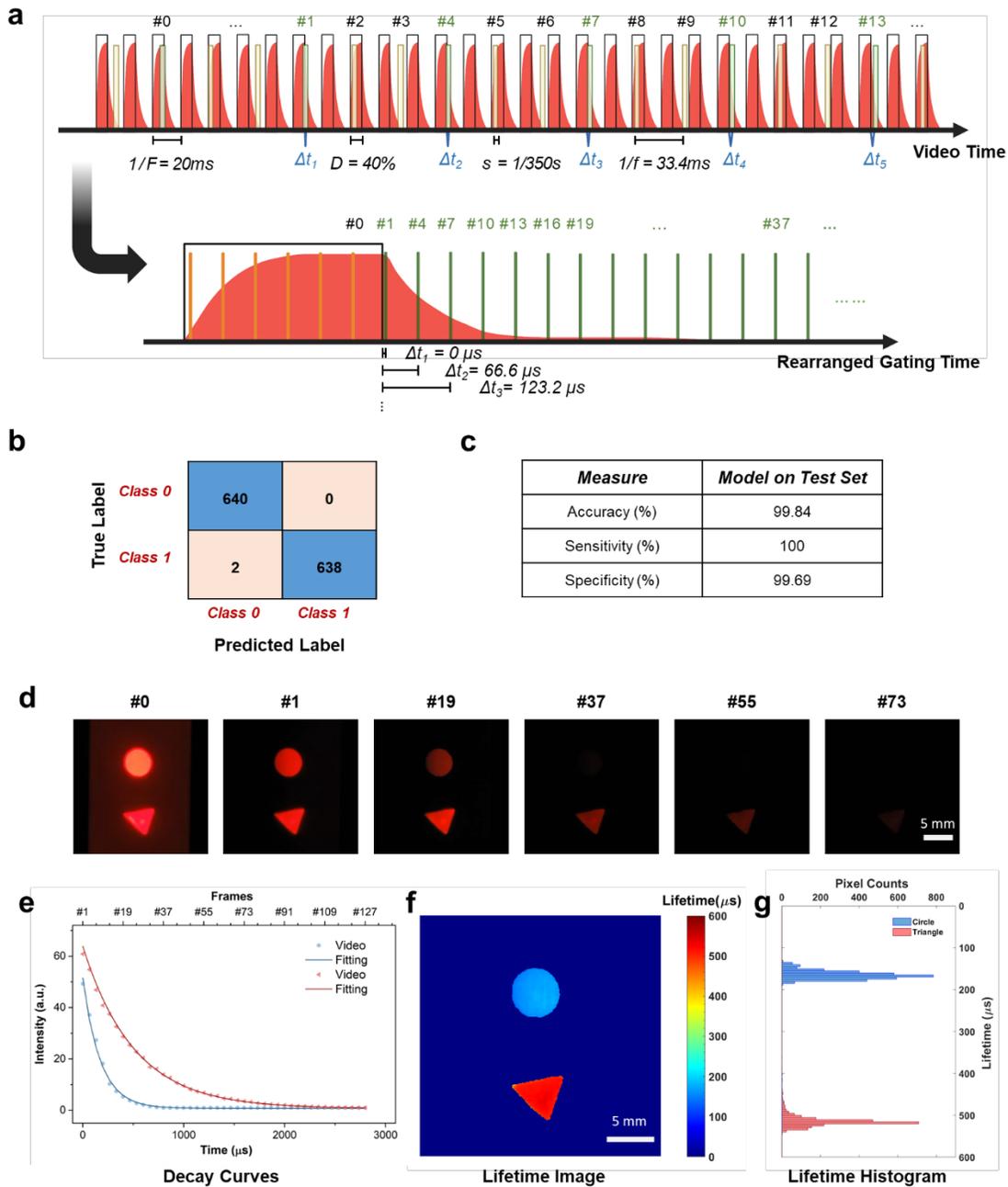

**Figure 5 | Demonstration of the smartphone V-chopper method for microsecond lifetime imaging. a,** Schematic of the smartphone video sequence and pulsed excitation timelines before and after frame extraction and gating images rearrangement. **b,** Confusion matrix of the CNN model. **c,** Performance of the CNN model trained by the combined training/validation datasets and applied on the test set. **d,** Representative #0 frame and gated frames extracted from the smartphone video by the CNN model. **e,** The decay curves of circle (blue) and triangle (red) over gated time; **f,** the reconstructed lifetime image of the microsecond probes, and **g,** the corresponding lifetime histograms from the two pattern spots.

To demonstrate the detection of sub-hundred microsecond lifetime on the smartphone V-chopper system, three different luminescent dyes were microprinted in a picture of a howling wolf (**Figure 6**). The wolf, ground, and moon consisted of Eu microbeads (lifetime ~500 µs), BHHCT-Eu chelate (lifetime ~250 µs), and tetracycline hydrochloride Eu dye (Tc-Eu) (lifetime <100 µs), respectively. The smartphone took videos at 30 fps (29.98 fps in real data set) with 1/500 s exposure time, and the UV LED was pulsed at a frequency of 30 Hz with a 40% duty cycle. That we are able to perform this experiment by using the same rate for both video recording (30 fps) and excitation pulse (30 Hz) is again due to the fact that the actual video rate (~29.98 fps) always deviates from the nominal value a little bit. Due to this mismatch, when a long video was recorded, the image frames were gradually shifting away from the UV pulses, generating gated UV-off images (**Figure 6a** and **Supporting Video V4**). Then, the gated frames in successive cycles were gathered to form the virtual decay image sequence. The delay time $\Delta t_n$ was measured as 22.2 µs, 44.4 µs, and 66.6 µs, respectively, for gated frame #1, #2, and #3, making it highly possible to resolve lifetime between 50-100 µs. This setting requires a long video being recorded to allow the frames scan over the whole decay curves, especially when the lifetime is over hundreds of microseconds. The minimum video duration (Equation 8) for efficient lifetime resolving is considered further in the **Discussion**.

**Figure 6b** displays frame #0 (UV on), frame #1 (the first gated frame after UV off), frame #60, and so on. Comparing the gated frames #1-160, it is obvious that the phosphor in the wolf is the most long lived, and the phosphor in the moon is the shortest lived—it is barely visible after frame #3. The decay curves of three dyes extracted from all gated frames were plotted in **Figure 6c** (green dots), and fitted by exponential functions (solid lines) to resolve the lifetimes. The recovered lifetime image (**Figure 6d**) shows three distinct lifetime values (510.4 µs, and 169.3 µs, and ~78.3 µs,) for the wolf, ground, and moon, respectively, corresponding to the advertised lifetime of each luminescent material (**Figure S2 and Table S2**). The data suggest that by applying the V-chopper concept, short lifetimes below 100 µs can be resolved on a portable smartphone reader device with a relatively low frame rate at 30 fps.

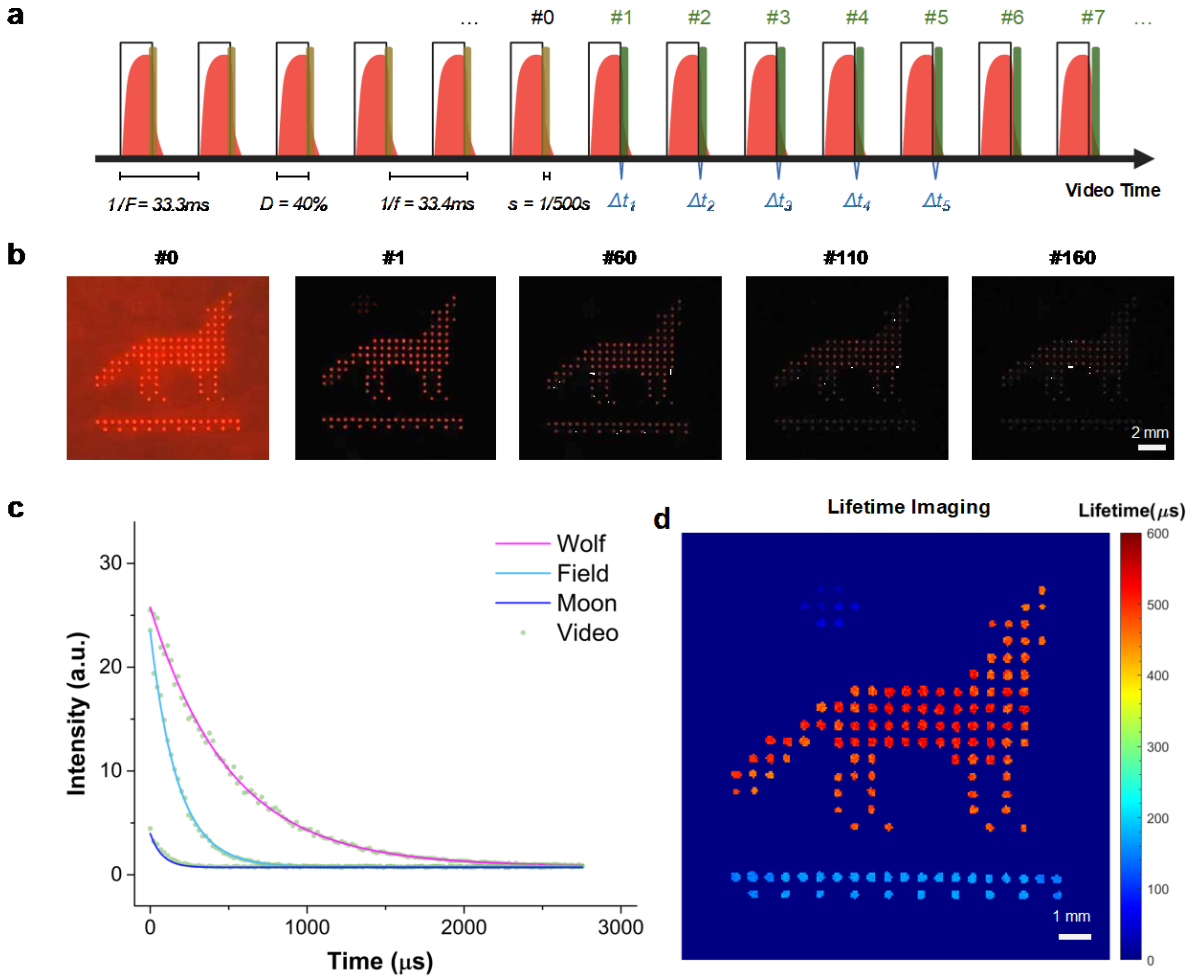

**Figure 6 | Detection of sub-hundred microsecond lifetime on the smartphone V-chopper device. a,** Schematic of the smartphone video sequence and pulsed excitation settings. **b,** Representative gated frames extracted from the smartphone video by the CNN algorithm. **c,** The decay curves of luminescent intensities over gated time. Green dots are actual experimental data, and solid lines are exponential fitting curves. **d,** The reconstructed lifetime image of the time-encoded pattern. The shortest lifetime pattern (moon) is clearly visualized.

## Discussion

A conventional time-resolved luminescence detection system is usually composed of a light source capable of pulsed excitation, a gated optical detector to capture time-dependent luminescence, and a synchronized control unit to provide the phase difference (or time delay) between the excitation and detection windows. Nowadays, the laser diodes and LEDs can achieve terrific temporal resolutions with repetition rates up to 100 MHz and pulse width of ns-µs, while maintaining the cost effectiveness and portability. Therefore, the gated optical detector becomes a key challenge in the implementation of time-gated detection, especially for a portable time-resolving system which needs a very high acquisition rate for multiple samplings in each cycle to fit the luminescence decay exponentially.

On the other hand, over the past decades, the smartphone has been widely explored as a capable analytical sensing and imaging platform in many POC applications, such as disease diagnostics, environmental monitoring, and screening for food contamination.[40–45] In most of the previous applications, the smartphone cameras are used to take individual photos for data analysis. Recently, with the rapid technological advancement of complementary metal oxide semiconductor (CMOS) cameras equipped on the smartphone, the users can have more control of the video capturing mode such as the exposure time, frame rate, and focusing distance, etc. In the recent half decade, many smartphone models on the market have achieved super high frame rate videos (above 100 fps) with high resolution definition and opened a new analytical method defined as smartphone videoscopy.[30]

However, previously reported smartphone-based lifetime quantification methods still require complicated mechanical chopper systems to achieve high temporal resolution, limiting the time-resolved technologies for POC use.[35–37] Here, instead of using expensive high-speed image sensors or complicated mechanical choppers, we demonstrated a virtual modulation method, the V-chopper, by generating controlled time shift between video frames and light pulses to reconstruct luminescence decay curves from multiple modulation cycles. The gated image extraction and rearrangement is fully automated by a CNN deep learning model. By applying the V-chopper method, the lifetime as short as few tens of microseconds can be resolved on a consumer smartphone device using a low video frame rate (e.g., 30 fps). Compared with previous mechanical chopper-based systems, the smartphone V-chopper platform is much more cost-effective and easier to implement. Meanwhile, the concept of the V-chopper can be broadly applied not only on smartphone detectors but also conventional digital cameras or image sensors. The latter can provide more precise control of frame rate and exposure time and has the potential to even resolve nanosecond lifetime.

To perform smartphone V-chopper, it is important to select a good combination of excitation and video recording settings. A Matlab-based simulation program (**Figure S3** and codes available at https://github.com/VictoriaYanWang/Smartphone-Lifetime-Imaging) was developed to study the interplay of the light pulse frequency ($F$), duty cycle ($D$), video recording rate ($f$), and shutter speed ($s$). The following general rules should be followed for a successful V-chopper implementation on the smartphone.

***Excitation Pulse.*** To detect the luminescence lifetime ($\tau$) precisely, the requirement for the LED

pulses is that the LED off time (defined by $(1 - D)/F$) should be longer than the whole decay curve, or longer than at least 3 times of lifetime $\tau$. In other words, the new excitation should not be turned on until the previous decay curve is completed.

$$(1 - D)/F \geq 3 \cdot \tau \qquad \text{(Equation 1)}$$

For instance, if the LED duty cycle is set at 40%, then the LED pulse frequency ($F$) should meet the following condition:

$$F \leq \frac{1}{5\tau} \qquad \text{(Equation 2)}$$

Therefore, based on the estimated lifetime of the dye to be detected, the repetition frequency of LED can also be selected accordingly. For example, for dyes with 10-millisecond lifetime, excitation with 20 Hz or lower frequencies (40% duty) should be used (**Table S3**). In contrast, for ultra-long lifetimes, excitation with 1 Hz or lower should be applied to match with the long decay curve.

***Shutter speed.*** Conditions for a successful time-resolved detection also involve the setting of shutter speed ($s$), which should be shorter than the LED off time, otherwise the camera will always capture LED on images. The condition can be expressed as:

$$s < (1 - D)/F \qquad \text{(Equation 3)}$$

For example, for LED frequency of 20 Hz, and 40% duty cycle, the shutter speed should be shorter than 0.03 or 1/33 seconds (**Tables S3**).

***Frame rate.*** The choice of frame rate on most smartphones is very limited. The values are discrete instead of being successively adjustable. However, the few options of preset frame rate will not limit the capacity of smartphone V-chopper device to resolve a broad range of lifetimes, including the short lifetimes in the microseconds. It is worth mentioning there is sometime a small drift to the preset frame rate when a video is taken on the smartphone. Based on the information stored in the video properties, the actual frame rate $f_{real}$ may be slightly different from the preset values (e.g., $f$=24, 30 or 60 fps) with a tiny drift $\sigma$ (0.01-0.30 Hz). For the current phone model we used, the $f_{real}$ is often 30.02 or 29.98 fps for a given preset rate of 30 fps.

$$f_{real} = f \pm \sigma \qquad \text{(Equation 4)}$$

The small drift of the smartphone frame rate provides a simple phase shifting mechanism between video frames and LED pulses, which provides the shortest delay time of gated frames that we can achieve in the current setup for measuring short luminescence lifetimes.

If the smartphone videoscopy is set at 30.00 fps sharp with 1/350 s exposure time, and LED has 50 Hz frequency with 40% duty cycle, the intensities of gated frames will be the same across the whole video since the frame window sits at the same delay time for each luminescence decay ($\Delta t_n$ is a constant, **Figure S4a**). However, for a frame rate of 29.98 fps (**Figure S4b**) or 30.02 fps (**Figure S4c**), the gated frames started to show modulated intensities across frames. That is because the small drift provides a phase shift to video frames which modulates $\Delta t_n$ over the decay curves. The varied $\Delta t_n$ allows video frames to sample different parts of the luminescence decay curves in multiple time-gated cycles. More specifically, if $f_{real} < f$, the intensity of the gated frames decreases as a function of time, meaning that gated frames scan the luminescence decay curve from high to low intensity (or shifting away from the LED pulses). In contrast, if $f_{real} > f$, the gated frames are shifting towards the LED pulses, so the intensity of the gated frames increases accordingly.

The delay time between successive frames $\Delta t$ will be equal to the drifting time between the actual frame rate and preset frame rate, which can be defined by

$$\Delta t = \left| \frac{1}{f} - \frac{1}{f_{real}} \right| = \frac{\sigma}{f \cdot f_{real}} \qquad \text{(Equation 5)}$$

Therefore the delay time between two successive gated frames $\Delta t_n$ will depend on frame numbers $m$ between each two, which means,

$$\Delta t_n = (n-1) \cdot m \cdot \Delta t \qquad \text{(Equation 6)}$$

And $m$ can be calculated by

$$m = f / \gcd(f, F) \qquad \text{(Equation 7)}$$

where gcd represents greatest common divisor. For example, for frame rate of 30 fps and UV pulse of 50 Hz (40% duty), $m = 3$; therefore $\Delta t_n = 3(n-1) \cdot \Delta t$, and $\Delta t_1$ is always close to 0. According to Equation 5, $\Delta t$ is 11 μs for $f = 30$ fps and $\sigma = 0.01$ fps, and $\Delta t$ will be 2.78 μs when $f = 60$ fps and same $\sigma$, It is clear that the smaller frame drift and higher frame rate that the smartphone camera can provide, the smaller time delay $\Delta t$ would be. The minimum delay time can predict the limit of detection (LOD) for lifetime by the smartphone V-chopper device on the order of tens of microseconds, which is equivalent to the results obtained by the previous smartphone systems equipped with mechanical choppers and motor for lifetime detection.[35–37]

Based on the above general rules and Equations 1-6, the recommended settings for successful smartphone V-chopper implementation on the different lifetime ranges can be found in the **Table S3**.

*Video duration.* The minimum duration of a video clip ($T_{min}$) to capture in order to scan over a whole decay curve for lifetime detection follows the equation:

$$T_{min} = \frac{(1-D)/F}{\Delta t} \cdot \frac{1}{f_{real}} = \frac{(1-D)/F}{\sigma/f} = \frac{f}{\sigma \cdot F} \cdot (1-D) \qquad \text{(Equation 8)}$$

According to Equation 8, the minimum necessary length of a video recorded to resolve the lifetime is proportional to the video frame rate (*f*) and inversely proportional to the LED pulsed frequency (*F*). As such, a lower video frame rate combined with higher frequency LED pulses is more time-efficient for lifetime detection on the smartphone.

In summary, a low-cost smartphone-based lifetime imaging platform has been developed for time-gated detection and 2D lifetime imaging over a broad range of lifetime from microseconds to seconds. To probe low microsecond lifetime events, a V-chopper method was demonstrated by modulating the LED pulses and smartphone video frame rate accordingly. Coupled with machine learning for gated image extraction, the V-chopper method gives the opportunities to resolve fast luminescence decay events on a low frame rate image sensor. The minimum lifetime that can be detected by the smartphone V-chopper system is about 75 μs, which is comparable or even lower than that obtained from previous mechanical chopper-based smartphone systems. This V-chopper method decouples the traditional time-resolved detection from expensive and complicated instruments. The miniaturized smartphone V-chopper system exhibits huge potentiality in lifetime imaging for various applications such as point-of-care biosensing. The methodology can also be a universal method, which can be applied on benchtop sensors to resolve even faster fluorescence events in the future.

**Online content**

Any methods, additional references, Nature Portfolio reporting summaries, source data, extended data, supplementary information, acknowledgements, peer review information; details of author contributions and competing interests; and statements of data and code availability are available at https://doi.org/

## Methods

**Preparation of the smartphone V-chopper device**

The smartphone V-chopper lifetime imaging prototype device consists of a 3D-printed enclosure, a UV LED (365nm, M365L3, Thorlabs), a condenser lens (ACL2520U-A, Thorlabs), a UV-enhanced reflection mirror (PFSQ10-03-F01, Thorlabs) and a smartphone (Samsung Galaxy S9). The sample glass slides can be placed inside of the enclosure on the bottom. The UV LED is controlled by a LED driver (LEDD1B, Thorlabs) and pulsed via a square wave voltage source (DG1062Z, Rigol). The highly divergent emission from the UV LED was first collimated by the aspheric condenser lens (f=20.1 mm, NA=0.60) and then evenly projected on the glass slide by a tilted reflection mirror (**Figure 1a**). The tilted angle of the 1" × 1" UV-enhanced mirror is designed to be ~67.5 degree relative to the slide surface, so the LED can deliver a uniform illumination to the sample slide and meanwhile the illumination center is aligned with the field of view (35 × 63 mm$^2$) of the smartphone. When detecting luminescent signals from Europium complex dyes, a 615 nm band-pass filter (87-739, Edmund Optics) can be mounted in front of the phone camera to eliminate the excitation interference. The Galaxy S9 smartphone has manual control of video settings, e.g., ISO, focal length, shutter speed, video frame rate, and image resolutions. A 60 fps video frame rate was used for lifetime detection of ultra-long luminescent materials (seconds). For the measurement of microsecond lifetime targets, a normal video rate (30 fps) was used instead combined with the V-chopper principle.

**Preparation of Eu dyes**

The long lifetime luminescence probes are often lanthanide-based complexes and nanoparticles, with luminescence lifetimes typically in the range of 1 μs to 10 ms. Ultra-long or persistent lifetimes can last for a few seconds or even up to minutes. Here, europium probes with different labelled lifetimes in microseconds, milliseconds, and seconds range have been used for demonstration of lifetime imaging with V-chopper on the smartphone videoscopy. To demonstrate the concept of using V-chopper for time-resolving detection and lifetime imaging, the dyes described above have been coated on paper substrates, which have strong autofluorescence to present as background noise. The ultra-long lifetime powders (micron size particles) with different colors and glow durations (shown in **Table S2**, Techno Glow) have been selected to demonstrate resolving lifetimes over hundreds of milliseconds and up to seconds. The red glowing powder contains calcium sulfide and the other three powders have composition of Strontium Aluminate Europium Dysprosium (SrAl$_2$O$_4$:Eu$^{2+}$, Dy$^{3+}$). To prepare different patterns, sticker labels were cut into "N", "C", "S", and "U" shapes to trap glow powders evenly on the adhesive side. Then four letters with different colored powders were placed on an autofluorescent paper substrate which was then sandwiched with two glass slides. Excess tetracycline hydrochloride (Tc) powder (T2525, TCI) was dissolved in 0.1 M $Na_2CO_3$ butter (pH=8). Then, $EuCl_3$ solution was added to generate Tc-Eu chelate. Filter paper (09801B, Fisherbrand) was then soaked in the Tc-Eu dye suspension and measured immediately when wet. BHHTC-Eu (59752, Sigma-Aldrich) with bright emission under UV was dissolved in DMSO with a concentration of 0.01 M to coat the filter paper. The original suspension of 0.2 µm Eu chelate polystyrene beads (S9347, Thermo Fisher) was diluted by 100 times with MilliQ water and then spiked on the filter paper.

The Eu chelate dye Tc-Eu, BHHCT-Eu, and Eu microbeads in the **Table S2** have been calibrated with a commercial time-resolved spectrometer (LP920, Edinburgh Instruments). The emission spectra peak of the Eu chelates is at ~615 nm when excited at the UV 365 nm (**Figure S5**). The emission spectrum of paper substrate and substrate with Eu dye were demonstrated in **Figure S6**. The red spectra (solid and dash) were measured when UV was on, and the blue ones with a 10 µs delay time after UV exposure. It shows the paper substrate has strong autofluorescence under UV (red dash), which can be eliminated with a 10 µs gate time (blue dash). However, the long-lived Eu dye is still luminescent, peaking at ~615 nm (blue solid). The lifetimes of Eu dyes have been measured by the time-resolved spectrometer as shown in **Table S2** and **Figure S2**.

**Simulation of the V-chopper mechanism**

To visualize the time delay ($\Delta t$) in the V-chopper method and study the interplay of the light pulse frequency, duty cycle, video recording rate, and shutter speed, a simulation program with user interface was designed in Matlab. The program is now available to download on GitHub (https://github.com/VictoriaYanWang/Smartphone-Lifetime-Imaging), which will help users to find an optimal excitation and data acquisition setting for a given lifetime target. When the LED pulse frequency (Hz), duty cycle (%), smartphone frame rate (fps), shutter speed (s), and estimated lifetime of the target dye are input in the interface window, two tracks of pulses will be generated, namely the waveform of LED (**Figure S3b**, blue solid line) and waveform of smartphone frames (**Figure S3b**, green solid line). Each smartphone frame will be assigned with a frame number (e.g., #1, #2, #3, etc.), shown on top the video frames. In addition, the decay curves of luminescence (**Figure S3b**, red dash line) will also be simulated following each LED excitation pulse. The X-axis represents time (s) and Y-axis with arbitrary units represents intensity for luminescence. The example simulation result shown in **Figure S3b** demonstrates when LED has a repetition rate of 50 Hz and duty cycle of 40%, and the frame rate of smartphone video is 30.02 fps with 1/350 s shutter speed. The gated frames will be found only at where the decay curve (**Figure**

**S3b**, red dashed line) and smartphone frames (**Figure S3b**, green solid line) are overlapped, marked in magenta (**Figure S3b**).

The size of the magenta area corresponds to the amount of light being collected in this frame, which is proportional to the luminescence intensity of the gated frames. Therefore, the luminescent intensity of each frame can be calculated by accumulating the area of magenta segments and plotted over time (**Figure S7a**). Several exponential curves (Decay n, n+1, n+2…) can be synthesized in a 30 s observation window. Among these virtual decay curves, Decay n+2 is a complete curve that can be used for accurate lifetime calculation. **Figure S7b** shows the actual frames extracted from the video which was taken with the settings used in **Figure S3b.**

**The CNN Classification model**

The basic structure of the CNN model is illustrated in **Figure 4**. To combat overfitting, several techniques were applied simultaneously: (i) batch normalization was utilized after each convolution and the fully connected layers; (ii) dropout was employed after the fully connected layer; (iii) early stopping was used to stop the training process when the validation loss reaches its minima. The RMSprop optimizer with the learning rate of 0.0001 was adopted to compute the model weights and biases.

For training, the smartphone image dataset was prepared with the balanced classes (0 or 1), each composed of 3200 images. A 60/20/20 split was utilized to initiate training, validation, and test sets, respectively. The performance of the CNN models was assessed using accuracy, sensitivity, and specificity that could be calculated based on the following equations:

$$Accuracy = (TP + TN)/(TP + TN + FP + FN) \qquad \text{(Equation 9)}$$

$$Sensitivity = TP/(TP + FN) \qquad \text{(Equation 10)}$$

$$Specificity = TN/(TN + FP) \qquad \text{(Equation 11)}$$

where TP is the true positive (the number of class 0 images classified correctly), TN is the true negative (the number of class 1 images classified correctly), FP is the false positive (the number of class 1 images the model classifies incorrectly as class 0), and FN is the false negative (the number of class 0 images the model classifies incorrectly as class 1).

After training, a user-friendly code was written to apply the trained CNN model to classify unseen smartphone images. The python code for training the CNN model, the model application code, and the smartphone images used as the dataset to develop and evaluate the model are freely available on GitHub (https://github.com/VictoriaYanWang/Smartphone-Lifetime-Imaging).

## Acknowledgements

This work was supported by North Carolina State University in the start-up funds provided to QW. Some of the schematic illustrations in this work were created with BioRender.com.

## Author contributions

Project conception and visualization was performed by YW, QW, and FL. Experiments were conducted by YW and RP. Data analysis was performed by YW and SS. Construction and design of the smartphone V-chopper device was completed by YW. ED assisted with design and execution of benchtop time-resolved spectrometer experiments. ZH assisted with the setup and execution of microprinter experiments. YW conducted computational study of the V-chopper methods and Matlab simulation program. SS conducted computational study of the CNN machine learning model and Python program. YW, SS and QW drafted the manuscript, and all authors contributed to the revision of the manuscript.

## Competing interests statement

The authors declare that they have no competing financial interests.

**Additional information**

**Supplementary information** The online version contains supplementary material and videos available at xxx.

**Correspondence and requests for materials** should be addressed to Qingshan Wei.

# Supporting Information

**Low-Rate Smartphone Videoscopy for Microsecond Luminescence Lifetime Imaging with Machine Learning**

Yan Wang[1], Sina Sadeghi[1], Rajesh Paul[1], Zach Hetzler[1], Evgeny Danilov[2], Frances S. Ligler[3], Qingshan Wei[1]*

[1] *Department of Chemical and Biomolecular Engineering, North Carolina State University, Raleigh, NC, 27695 USA.*

[2] *Department of Chemistry, North Carolina State University, Raleigh, NC, 27695 USA.*

[3] *Department of Biomedical Engineering, Texas A&M University, College Station, TX 77843, USA.*

* Corresponding email: qwei3@ncsu.edu

# Contents

Supplementary Figures S1-S7

Supporting Tables S1-3

Supporting Videos V1-4 (Attached Individually)

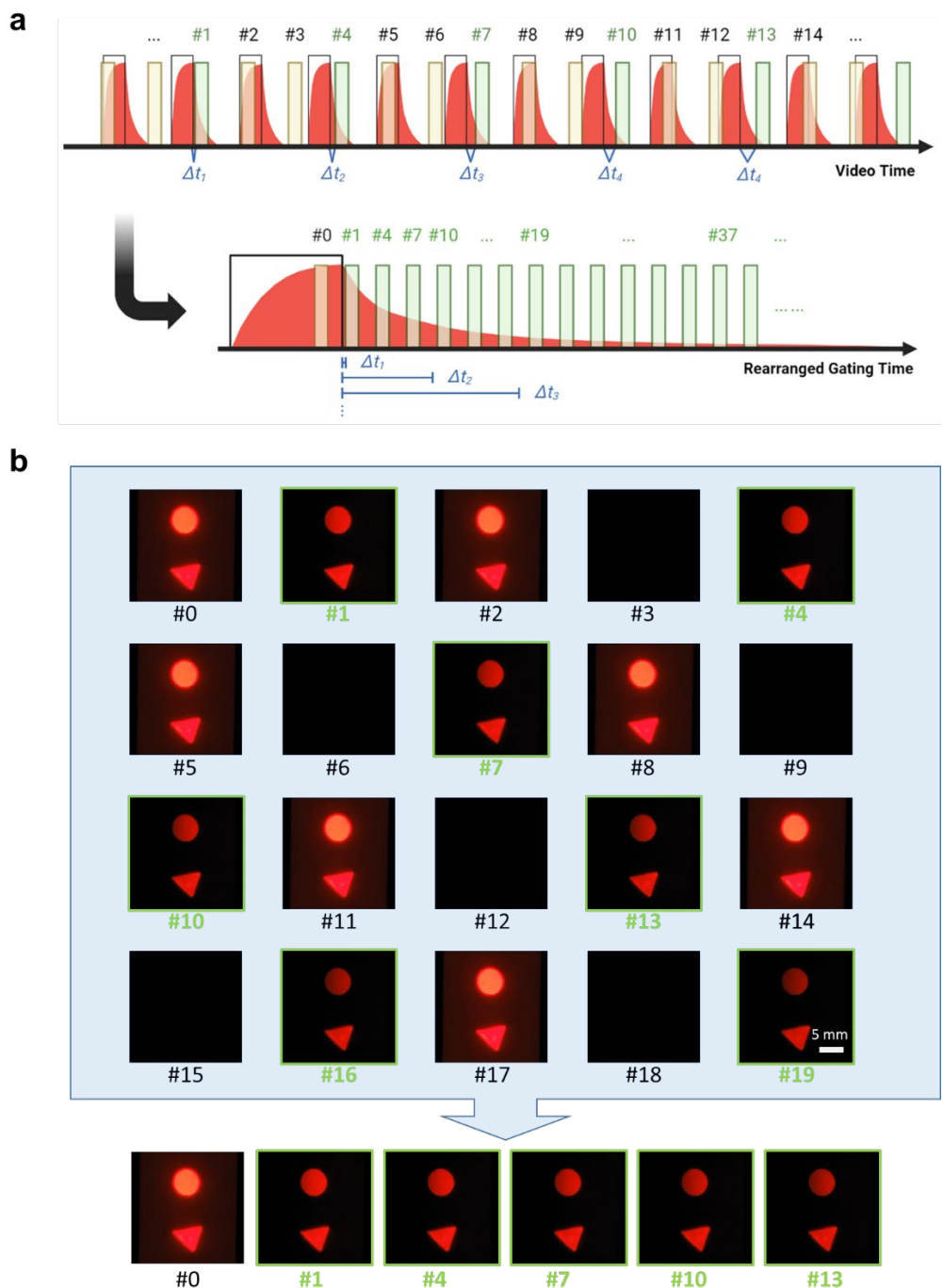

**Figure S1 | Raw video frames from the smartphone for microsecond lifetime imaging. a,** Schematic of the smartphone video sequence and pulsed excitation timelines before and after frame extraction and gated images rearrangement. In the upper row, the large clear bars (black edge) indicate the timing when the UV excitation was on, while the translucent bars indicate the timing of the detection measurements. The red curves reflect time-dependent emission intensity. In the lower row, data collected from frames (translucent

bars) without the excitation on were collected to measure the emission lifetimes. **b,** The gated images was classified from the original extracted frames by applying the CNN model.

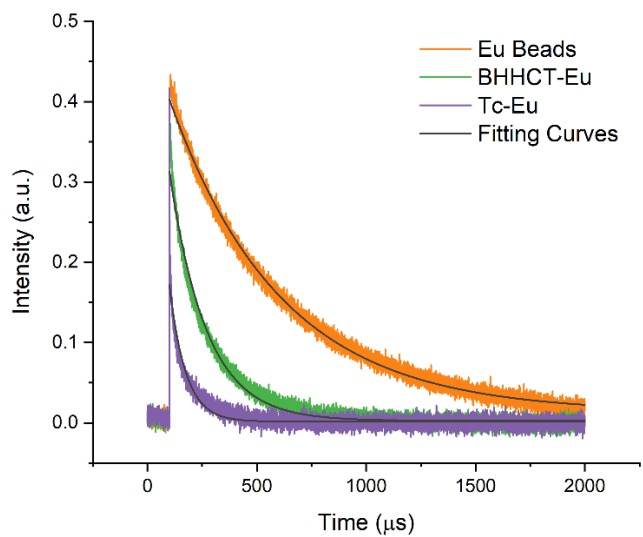

**Figure S2 | Calibrated luminescent decay curves of three Eu dyes used in this paper.** A commercial benchtop time-resolved spectrometer (LP920, Edinburgh Instruments) was used to calibrate the lifetimes, which have been demonstrated in **Table S2**.

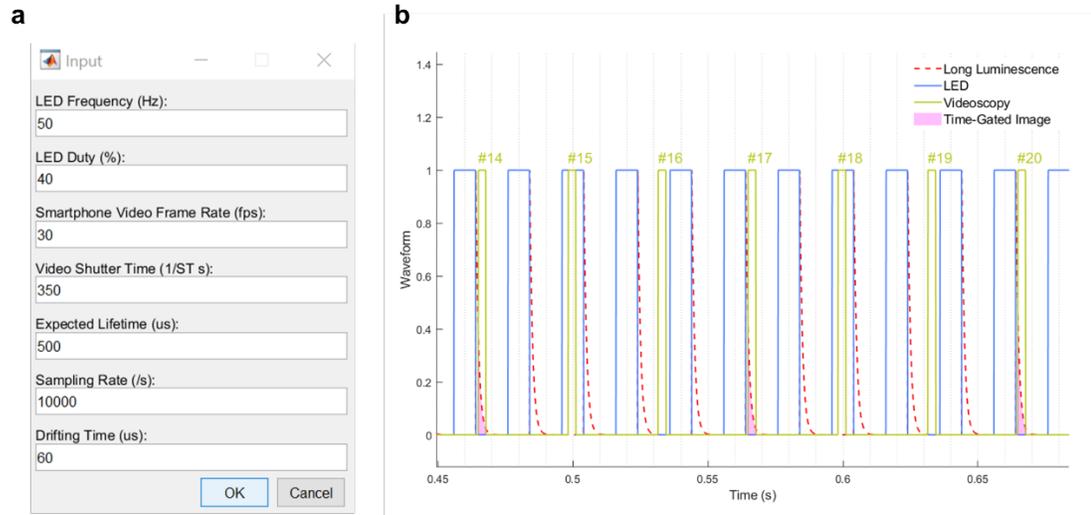

**Figure S3 | Matlab GUI for smartphone videoscopy simulation. a,** Input window. **b,** Signal train of simulated LED pulses (blue solid), V-chopper (green solid) and decay of long luminescence (red dash). The overlap area of V-chopper and luminescence represents the intensity of gated frames captured in the sequence (magenta segments), which have been plotted in **Figure S4**.

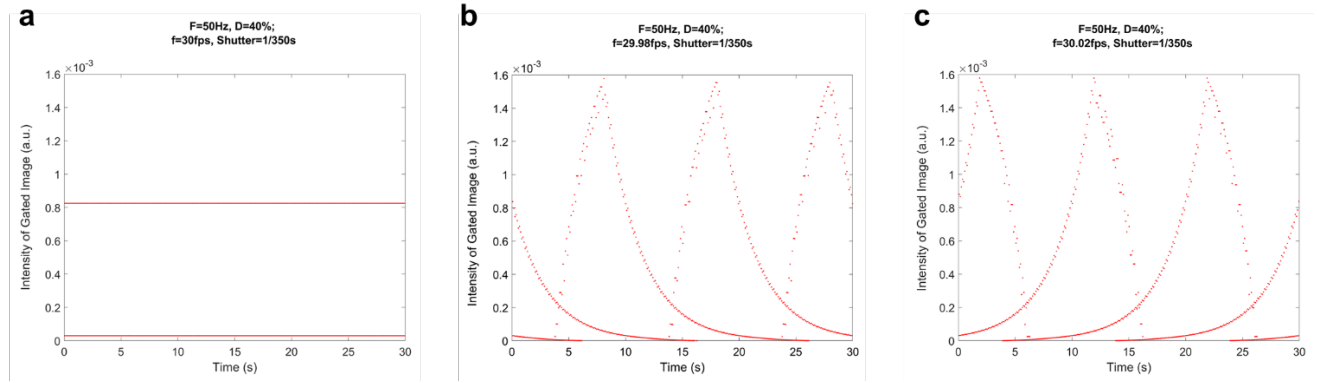

**Figure S4 | Simulated intensity of gating frames in a 30 seconds video when recorded with (b and c) or without (a) drifting to preset 30 fps frame rate. a,** The intensities of gated frames are the same across the whole video when frame rate is sharp 30 fps ($\Delta t_n$ is a constant). **b and c,** The gated frames started to show modulated intensities across frames, and **b** shows decreased intensity modulation when $f_{real} < f$, and **c** shows increased modulation if $f_{real} > f$.

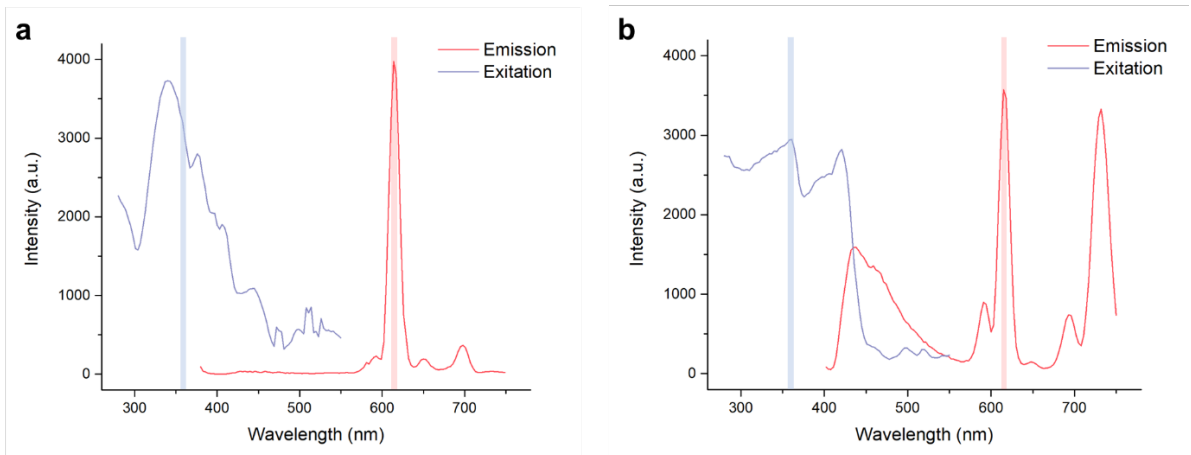

**Figure S5 | Excitation and emission spectra of Eu dyes. a,** Eu microbeads spectrum measured in a water suspension. **b,** Tc-Eu dye spectrum measured in 0.1 M $Na_2CO_3$ butter (pH=8).

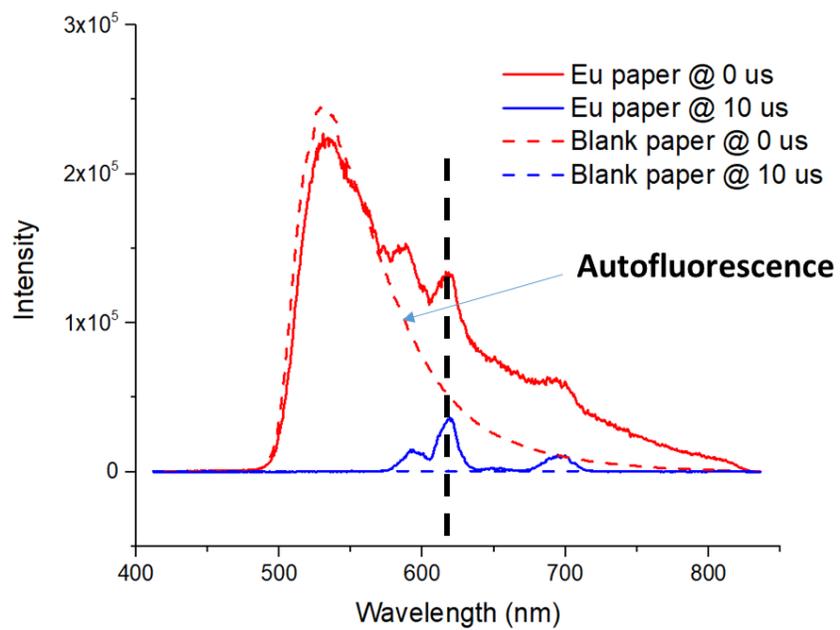

**Figure S6 | The emission spectra of blank paper substrate and substrate soaked with Eu dye.**

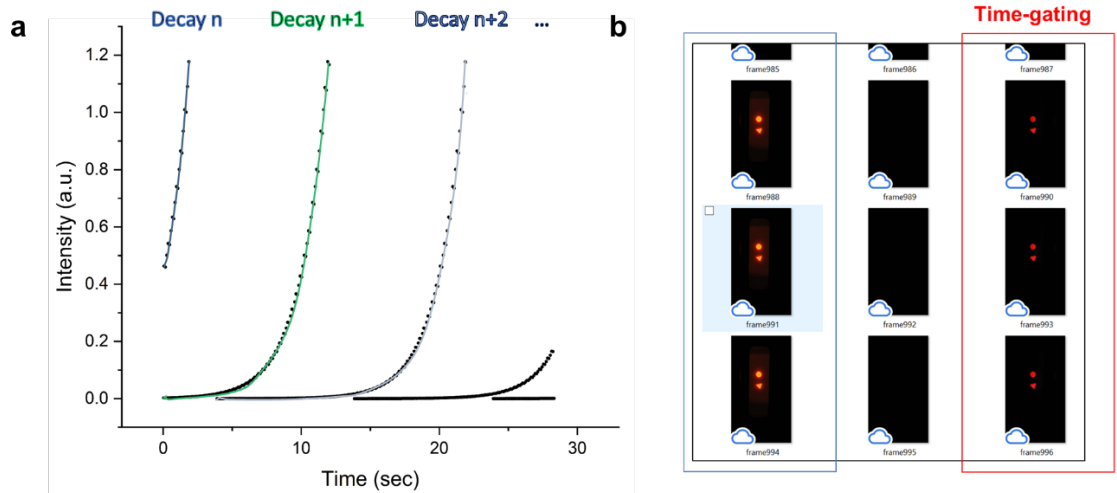

**Figure S7 | a,** Luminescence intensity of gated frames calculated based on the simulated results. In a 30 seconds video clip, the whole decay of long-lived luminescence signal has been fully captured in the Decay n+2. **b,** Actual gated frames that mixed with autofluorescent frames in the original video.

**Table S1 | The CNN model complexity in terms of number of learnable parameters.**

| Layer | Number of Learnable Parameters |
|---|---|
| Convolution 1 | 160 |
| Batch Normalization 1 | 32 |
| Max Pooling 1 | 0 |
| Convolution 2 | 8224 |
| Batch Normalization 2 | 64 |
| Max Pooling 2 | 0 |
| Convolution 3 | 18496 |
| Batch Normalization 3 | 128 |
| Max Pooling 3 | 0 |
| Fully Connected Layer | 160100 |
| Batch Normalization 4 | 200 |
| Output Layer | 202 |
| *Total Number of Learnable Parameters* | *187606* |

**Table S2 | List of Eu dyes for lifetime imaging used in this paper.**

| Probe Dye | Labelled Lifetime | Calibrated Lifetime | Luminescence Color | Form |
|---|---|---|---|---|
| Calcium Sulfide | 1-4 Hours | 1.5 s | Red | Micron Particles |
| Strontium Aluminate Europium Dysprosium | 4-8 Hours | 5 s | Malachite Green | Micron Particles |
| | 8-12 Hours | 7 s | Jade Green | Micron Particles |
| | Over 12 Hours | 31 s | Cyan | Micron Particles |
| Tc-Eu | 100 µs | 76.6 ± 2.7 µs | Red | Powder |
| BHHCT-Eu | 250 µs | 167.2 ± 1.4 µs | Red | Powder |
| Eu Chelate Beads | 500 µs | 516.4 ± 7.0 µs | Red | Polystyrene Beads |

**Table S3 | Recommended example settings of LED and V-chopper for different lifetime ranges on the smartphone.**

| Lifetime ($\tau$) | LED Frequency ($F$) (40% duty) | V-Chopper | |
|---|---|---|---|
| | | Frame Rate ($f$) | Shutter Speed ($s$) |
| 100 µs | 20 Hz | 30 fps | <1/500 s |
| 500 µs | 20 Hz | 30 fps | 1/350 s |
| 1 ms | 20 Hz | 30 fps | 1/350 s |
| 10 ms | 20 Hz | 30 fps | 1/350 s |
| 50 ms | 1 Hz | 60 fps | 1/350 s |
| 200 ms | 1 Hz | 60 fps | 1/350 s |
| 1s | $\leq 0.2$ Hz | 60 fps | 1/350 s |

**Video V1 | Raw video recorded by Galaxy S9 smartphone for the detection of ultra-long lifetime in the seconds range.** The video was recorded at 60 fps with exposure time of 1/60 s. After the UV excitation, the letters of "N", "C", "S", "U" emit glows for different time durations. The long luminescence decays are able to be captured and resolved by the 60 fps video frames directly in a single decay cycle. The video was played at a 4× speed to demonstrate the whole process of glowing from four letters.

**Video V2-3 | Raw videos (V2: 1× original speed, V3: 0.1× original speed) recorded by smartphone V-chopper device for lifetime measurement in the microsecond range.** The original video (V2) was recorded with a frame rate of 29.98 fps and exposure time of 1/350 s, meanwhile the LED was pulsed at 50 Hz with a 40% duty cycle. V2 was played at 1× speed, while V3 was played at a lower speed of 0.1× to clearly show the grated frames (zero background). These gated frames were then extracted by the CNN model as mentioned in the main text and used to resolve the lifetime image.

**Video V4 | Raw video (1× original speed) recorded by smartphone V-chopper device for multiplexed lifetime detection in the sub-hundred to hundreds of microseconds range.** The video (V4) was recorded with a frame rate of 29.98 fps and exposure time of 1/500 s, meanwhile the LED was pulsed at 30 Hz with a 40% duty cycle. The video started with LED-on frames and followed by sequential gated images from multiple decay cycles. As illustrated in **Figure 6a**, after the LED off, a series of gated frames (one gated image per cycle, m = 1) with increasing time delay $\Delta t_n$ were captured. The "Moon" (sub-hundred lifetime) disappeared very fast in two or three frames, but the "Wolf" with the longest lifetime is lasting much longer.